\documentclass[10pt,letterpaper]{article}
\usepackage[top=0.85in,left=2.75in,footskip=0.75in,marginparwidth=2in]{geometry}

\usepackage[utf8]{inputenc}

\usepackage{cite}
\usepackage{multirow}
\usepackage{multicol}
\usepackage{makecell}
\usepackage{threeparttable}
\usepackage{ragged2e}
\DeclareUnicodeCharacter{2212}{-}
\usepackage{nameref,hyperref}

\usepackage[right]{lineno}

\usepackage{microtype}
\DisableLigatures[f]{encoding = *, family = * }

\raggedright
\usepackage{changepage}

\usepackage[aboveskip=1pt,labelfont=bf,labelsep=period,singlelinecheck=off]{caption}

\makeatletter
\renewcommand{\@biblabel}[1]{\quad#1.}
\makeatother

\usepackage{lastpage,fancyhdr,graphicx}
\usepackage{epstopdf}
\fancyheadoffset[L]{2.25in}
\fancyfootoffset[L]{2.25in}

\usepackage{color}

\definecolor{Gray}{gray}{.25}

\usepackage{graphicx}

\usepackage{sidecap}

\usepackage{wrapfig}
\usepackage[pscoord]{eso-pic}
\usepackage[fulladjust]{marginnote}
\reversemarginpar
\usepackage{graphicx}
\graphicspath{{figures/}}
\usepackage{subcaption}
\usepackage{mwe}
\begin{document}
\vspace*{0.35in}

% title goes here:
\begin{flushleft}
{\Large
\textbf\newline{Detection of Maternal and Fetal Stress from the Electrocardiogram with Self-Supervised Representation Learning}
}
\newline
% authors go here:
\\
Pritam Sarkar\textsuperscript{1,\textdagger},
Silvia Lobmaier\textsuperscript{2,*,\textdagger},
Bibiana Fabre\textsuperscript{5},
Diego González\textsuperscript{5},
Alexander Mueller\textsuperscript{6},
Martin G. Frasch\textsuperscript{3,4,*},
Marta C. Antonelli\textsuperscript{2,7,*}
Ali Etemad\textsuperscript{1,*}
\\
\bigskip
\bf{1} Dept. of ECE \& Ingenuity Labs Research Institute, Queen’s University, Kingston, Ontario, Canada
\\
\bf{2} Dept. of Obstetrics and Gynecology, Technical University of Munich, Munich, Germany
\\
\bf{3} Dept. of Obstetrics and Gynaecology, University of Washington, Seattle, Washington, USA
\\
\bf{4} Center on Human Development and Disability, University of Washington, Seattle, Washington, USA
\\
\bf{5} Facultad de Farmacia y Bioquímica, Universidad de Buenos Aires, Buenos Aires, Argentina
\\
\bf{6} Dept. of Cardiology, Technical University of Munich, Munich, Germany
\\
\bf{7} IBCN, Facultad de Medicina, Universidad de Buenos Aires, Buenos Aires, Argentina
\\
\bigskip
\textdagger  Co-first authors

*Co-corresponding authors
% \footnote{ali.etemad@queensu.ca; mfrasch@uw.edu; marta.antonelli@mri.tum.de; silvia.lobmaier@tum.de}

\end{flushleft}
\justifying
\section*{Abstract}
In the pregnant mother and her fetus, chronic prenatal stress results in entrainment of the fetal heartbeat by the maternal heartbeat, quantified by the fetal stress index (FSI). Deep learning (DL) is capable of pattern detection in complex medical data with high accuracy in noisy real-life environments, but little is known about DL's utility in non-invasive biometric monitoring during pregnancy. A recently established self-supervised learning (SSL) approach to DL provides emotional recognition from electrocardiogram (ECG). We hypothesized that SSL will identify chronically stressed mother-fetus dyads from the raw maternal abdominal electrocardiograms (aECG), containing fetal and maternal ECG. Chronically stressed mothers and controls matched at enrolment at 32 weeks of gestation were studied. We validated the chronic stress exposure by psychological inventory, maternal hair cortisol and FSI. We tested two variants of SSL architecture, one trained on the generic ECG features for emotional recognition obtained from public datasets and another transfer-learned on a subset of our data. Our DL models accurately detect the chronic stress exposure group (AUROC=0.982$\pm$0.002), the individual psychological stress score (R2=0.943$\pm$0.009) and FSI at 34 weeks of gestation (R2=0.946$\pm$0.013), as well as the maternal hair cortisol at birth reflecting chronic stress exposure (0.931$\pm$0.006). The best performance was achieved with the DL model trained on the public dataset and using maternal ECG alone. The present DL approach provides a novel source of physiological insights into complex multi-modal relationships between different regulatory systems exposed to chronic stress. The final DL model can be deployed in low-cost regular ECG biosensors as a simple, ubiquitous early stress detection and monitoring tool during pregnancy. This discovery should enable early behavioral interventions. 

% the * after section prevents numbering
\section{Introduction}
Maternal chronic stress during pregnancy programs the fetal brain for altered developmental trajectories. We showed that in stressed mother-fetus dyads, this results in measurable synchronization of the fetal heartbeat by the maternal heartbeat, quantified by the fetal stress index (FSI) \cite{lobmaier2020fetal}. Can this biophysical phenomenon be scaled to an easily deployable biomarker of chronic stress in pregnant mothers to help guide early interventions which can reverse altered fetal developmental trajectories? 

Deep learning (DL)-based approaches \cite{lecun2015deep} to pattern detection in complex physiological data have shown high accuracy in noisy real-life environments \cite{sarkar2019classification,ross2019toward}. Nonetheless, little is known about their utility in the setting of non-invasive biometrics obtained during human pregnancy. 

\begin{figure}[ht]
    \includegraphics[width=\textwidth]{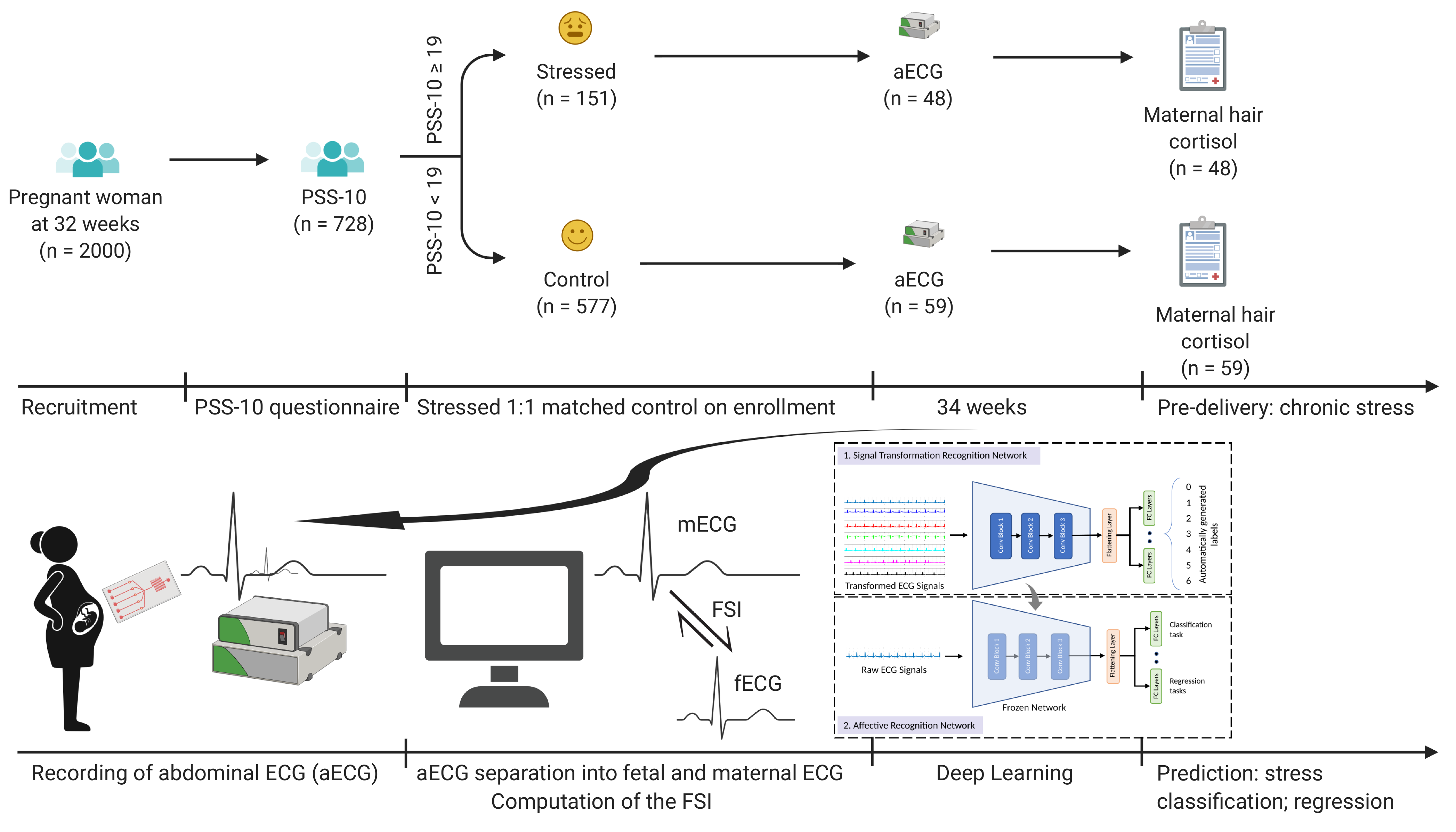}
    \caption{Summary of the approach: Prenatal Distress Questionnaire (PDQ) and Prenatal Stress Score (PSS-10) were determined in 32 weeks pregnant women classifying them as stressed group or matched controls. At 34 weeks, abdominal ECG (aECG) was recorded and prior to delivery, maternal hair was sampled for cortisol measurements reflecting chronic stress exposure over the past two months. The aECG was deconvoluted into fetal and maternal ECG (fECG, mECG) from which Fetal Stress Index (FSI) was computed, reflecting joint maternal and fetal chronic stress exposure. Deep Learning using a self-supervised learning framework ensued on aECG and mECG (fECG did not qualify due to signal quality) to detect stress group status (i.e., classification) and values of cortisol, FSI, PDQ, and PSS-10 (i.e., regression).}
    \label{fancy_fig}
\end{figure}

Here, we hypothesized that a DL approach to pattern recognition in maternal abdominal electrocardiograms (aECG) obtained in chronically stressed mothers and controls matched at enrolment at 32 weeks of gestation will detect chronic stress in mother-fetus dyads, i.e., a DL classification model (Figure \ref{fancy_fig}). 

We validated the exposure to stress by psychological inventory, molecular and biophysical biomarkers including maternal hair cortisol and FSI, respectively. Then, we tested the correlation between these exposure measures and the aECG and maternal ECG (mECG) features captured by the DL pipeline, i.e., DL regression model. We implemented the DL pipeline using the recently established self-supervised learning (SSL) approach that provides emotional recognition from ECG \cite{sarkar2020selficassp,sarkar2020selftaffc}.

We tested two variants of SSL architecture, one trained on the generic ECG features for emotion recognition obtained from public datasets and another transfer-learned on a subset of the composite aECG (which includes fetal ECG, fECG) or mECG data. Our studies of the model’s performance in regression tasks and with or without the inclusion of the fetal ECG signal reveal a rich structure correlating to psychological, molecular, and biophysical biomarkers of maternal and fetal stress exposure at 34 weeks of gestation and at birth.

\section{Results}

\subsection{Differences in the datasets}
There were no differences in age between the cohorts of our and the public datasets and the total number of subjects; in the public dataset used for training there were 103 subjects compared to 107 in FELICITy dataset (Table \ref{tab:tab1}). 
\begin{table}[!b]
% \begin{adjustwidth}{-1.5in}{0in} % comment out/remove adjustwidth environment if table fits in text column.
% \centering
\setlength\tabcolsep{4pt}
\caption{Demographic and dataset characteristics.}
\begin{tabular}{llllll}
\hline
% \multicolumn{4}{|l|}{\bf Heading 1} & \multicolumn{3}{|l|}{\bf Heading 2}\\ \hline
Dataset & AMIGOS & DREAMER & WESAD & SWELL & FELICITy\\
\hline\hline
No. of Participants & 40 & 23 & 15 & 25 & 107\\
Female/Male & 13/27 & 9/14& 3/12 & 8/17 & 107/0\\
Age & 28.3 (21-40) & 26.6$\pm$2.7 & 27.5$\pm$2.4 & 25$\pm$3.25  & 33$\pm$4\\
Duration (min.) & 95 & 60 & 120 & 95 & 46\\
Sampling rate (\textit{Hz}) & 256 & 256 & 700 & 2048 & 900\\
\hline
\end{tabular}
\label{tab:tab1}
% \end{adjustwidth}
\end{table} 

A clear difference existed in the gender composition, albeit its impact on the model performance remains uncertain. ECG duration was more variable in the public dataset than in the FELICITy dataset and so was the sampling rate. However, it remains also uncertain whether this had any impact on the model performance, especially since all ECG was resampled at 256 \textit{Hz} for the DL pipeline. It is possible that such variance in data quality and the composition of the participants made the model more robust, but this conjecture would need to be tested in future work.

We compared the model performance for detecting stressed mother-fetus dyads as well as predicting maternal hair cortisol, FSI, PDQ, and PSS values depending on two factors: the source of ECG (aECG, mECG) and the source of the trained model (learning from the FELICITy dataset - the first SSL approach, or transfer-learning from the public datasets - the second SSL approach) (Tables \ref{tab:tab2},  \ref{tab:tab3}).

\begin{table}[!ht]
% \begin{adjustwidth}{-1.5in}{0in} % comment out/remove adjustwidth environment if table fits in text column.
% \centering
% \fontsize{9pt}{10pt}\selectfont
\setlength\tabcolsep{4pt}
\caption{Detection of stressed mothers by self-supervised learning trained on the FELICITy and public datasets.}
\label{tab:tab2}
\begin{threeparttable}
\begin{tabular}{llllllll}
\hline
\multicolumn{8}{c}{\textbf{FELICITy dataset}}\\ \hline
Source & Accuracy & F1 Score & Sensitivity & Specificity & PPV & NPV & AUROC\\ \hline
aECG & $\makecell[l]{0.795 \pm\\ 0.023}$ & $\makecell[l]{0.777 \pm\\ 0.022}$ & $\makecell[l]{0.779 \pm\\ 0.031}$ & $\makecell[l]{0.809 \pm\\ 0.045}$ & $\makecell[l]{0.777 \pm\\ 0.039}$ & $\makecell[l]{0.812 \pm\\ 0.020}$ & $\makecell[l]{0.794 \pm\\ 0.022}$\\\hline
mECG & $\makecell[l]{0.931 \pm\\ 0.093}$ & $\makecell[l]{0.925 \pm\\ 0.101}$ & $\makecell[l]{0.924 \pm\\  0.101}$ & $\makecell[l]{0.937 \pm\\ 0.087}$ & $\makecell[l]{0.926 \pm\\ 0.102}$ & $\makecell[l]{0.936 \pm\\ 0.086}$ & $\makecell[l]{0.931 \pm\\ 0.094}$\\
\hline

\hline
\multicolumn{8}{c}{\textbf{Public datasets}}\\ \hline
Source & Accuracy & F1 Score & Sensitivity & Specificity & PPV & NPV & AUROC\\ \hline
aECG & $\makecell[l]{0.936 \pm\\ 0.002}$ & $\makecell[l]{0.930 \pm\\ 0.003^*}$ & $\makecell[l]{0.926 \pm\\ 0.008^*}$ & $\makecell[l]{0.945 \pm\\ 0.004^*}$ & $\makecell[l]{0.935 \pm\\ 0.004^*}$ & $\makecell[l]{0.938 \pm\\ 0.006^*}$ & $\makecell[l]{0.936 \pm\\ 0.002^*}$\\\hline
mECG & $\makecell[l]{0.982 \pm\\ 0.003}$ & $\makecell[l]{0.980 \pm\\ 0.003^{\#*}}$ & $\makecell[l]{0.982 \pm\\ 0.004^{\#*}}$ & $\makecell[l]{0.982 \pm\\ 0.006^{\#*}}$ & $\makecell[l]{0.979 \pm\\ 0.007^\#}$ & $\makecell[l]{0.985 \pm\\ 0.003^\#}$ & $\makecell[l]{0.982 \pm\\ 0.002^{\#*}}$\\
\hline
\end{tabular}

% \end{adjustwidth}
\begin{tablenotes}
\item[$^*$] Public versus FELICITy dataset, Mann Whitney U test.
\item[$^\#$] mECG versus aECG within the same dataset, Mann Whitney U test.
% \item Significance assumed if p $<$ $0.025$ accounting for two comparisons (using Bonferroni-Holm correction).
\item[Statistical significance at p $<$ $0.025$ accounting for two comparisons (using Bonferroni-Holm correction).]
\end{tablenotes}
\end{threeparttable}
\end{table}

\subsection{Identification of stressed mother-fetus dyads: DL classification task}
Within the FELICITy dataset, the ECG source made no difference, but using the public dataset improved the F1 score, sensitivity, specificity and AUROC regardless of the ECG source (Table \ref{tab:tab2}). In comparison to the FELICITy dataset, training on the public dataset while using aECG improved performance across all metrics except the accuracy. Accuracy was excellent overall and stood out as not being influenced by the ECG source or the origin of the trained model. Again in comparison to the FELICITy dataset, training on the public dataset while using mECG also improved performance overall, except accuracy, PPV, and NPV. This was because mECG in general boosted the performance regardless of how the model was trained - on the FELICITy or the public datasets.
The best group classification performance overall, across all metrics, was achieved using the public dataset and mECG.

\subsection{Prediction of stress biomarkers: DL regression task}

Recognizing the spread of PSS-10 scores, in the present study we also assessed the regression relationship between the scores and emotional recognition performance in our DL model (Table \ref{tab:tab3}). The model performance results were similar for the regression analyses. We see overall similar improvements and best performance for all biomarkers when using mECG and the public dataset. When using the model trained on the FELICITy dataset, there was no difference in prediction for all biomarkers when using aECG or mECG. This suggests there is enough information in the mECG and the model trained on the FELICITy dataset. In contrast, using the model trained on the public dataset improved the performance regardless of the source of data, aECG or mECG. 

For aECG on the FELICITy dataset, the model performed poorly for all biomarkers. Using mECG instead brought no significant improvement. When training on the public dataset, the performance improved on both aECG and mECG for cortisol, FSI, and PDQ, but not for PSS when using mECG, because it is already quite accurate when trained on the FELICITy dataset. 
In other words, the prediction of the PSS scores achieves highest performance when using the SSL pipeline trained on the FELICITy dataset and using mECG rather than the composite aECG, i.e., a signal containing maternal and fetal ECG combined.

\begin{table}[!ht]
\caption{Prediction of biomarkers by self-supervised learning on the FELICITy and public datasets.}
\label{tab:tab3}
\begin{threeparttable}
\begin{tabular}{llll}
\hline
% \multirow{2}{l}{Heading 1}
Task & Source & R2 FELICITy dataset & R2 Public datasets\\\hline\hline
\multirow{2}{*}{Cortisol}
 & aECG & $0.456 \pm 0.053$ & $0.801 \pm 0.009^*$\\ 
 & mECG & $0.743 \pm 0.322$ & $0.931 \pm 0.006^{\#*}$ \\\hline
\multirow{2}{*}{FSI}
 & aECG & $0.362 \pm  0.052$ & $0.768 \pm 0.018^*$\\
 & mECG & $0.780 \pm 0.274$ & $0.946 \pm 0.013^{\#*}$\\\hline
\multirow{2}{*}{PDQ}
 & aECG & $0.408 \pm 0.062$ & $0.781 \pm 0.019^*$\\
 & mECG & $0.789 \pm 0.302$ & $0.961 \pm 0.010^{\#*}$\\\hline
\multirow{2}{*}{PSS}
 & aECG & $0.344 \pm 0.072$ & $0.761 \pm 0.012^*$\\
 & mECG & $0.780 \pm 0.294$ & $0.943 \pm 0.009^\#$\\\hline
\end{tabular}
\begin{tablenotes}
\item[$^*$] Public versus FELICITy dataset, Mann Whitney U test
\item[$^\#$] mECG versus aECG within the same dataset, Mann Whitney U test
\item[Statistical significance at p $<$ $0.025$ accounting for two comparisons (using Bonferroni-Holm correction).]
\end{tablenotes}
\end{threeparttable}
\end{table}

For FSI and PDQ, it appears that the effect of the regression improvement by using the public dataset is dependent not on the biomarker, but on the data source, i.e., aECG versus mECG. This may be explained by the richer intrinsic structure of aECG compared to the uniquely maternal sourced mECG, which is better captured by the public dataset. The public dataset was also richer than the FELICITy dataset with regard to participants’ gender composition, ECG sampling rate and duration (Table \ref{tab:tab1}).  

\begin{figure}[ht]
    \includegraphics[width=0.6\textwidth]{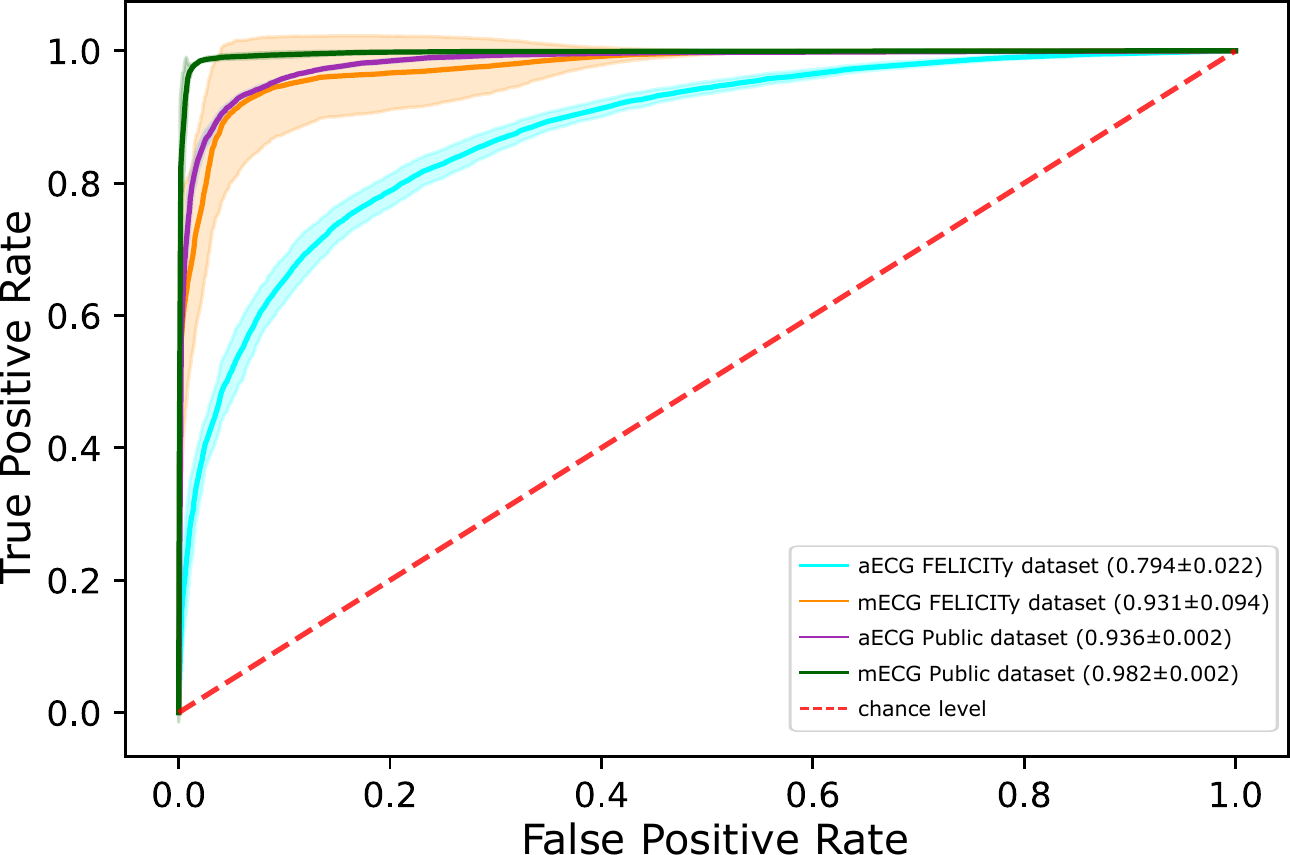}
    \caption{AUROC of SSL models trained on the public and FELICITy datasets to identify stressed and non-stressed mother-fetus dyads from aECG or mECG. Mean AUROC values are marked as solid lines and standard deviations across 5-folds are marked as shaded regions.}
    \label{roc}
\end{figure}

Overall, using the raw aECG decreases the model performance on both classification and regression. Identification of the effects of chronic stress and a highly accurate prediction of its effects on cortisol, FSI, PDQ, and PSS is possible from maternal ECG alone using the SSL model trained on the public dataset and using FELICITy dataset does not improve this performance neither for classification nor for regression. This is visualized in Figure \ref{roc}.

\section{Discussion}

Chronic stress is one of the most common modifiers of fetal and postnatal development with lifelong lasting effects on health and no systematic prevention programs exist today \cite{frasch2018non,desplats2019microglial}. The present findings provide a solution. First, confirming our hypothesis, we report a scalable and readily deployable approach using an SSL model of DL to identify chronically stressed mother-fetus dyads and predict their biochemical, biophysical, and psychological characteristics from a regular mECG with a high degree of accuracy. The excellent performance of the model trained on the public dataset suggests a high probability of generalizability of our findings to new data.

\begin{figure}
    \centering
    \includegraphics[width=0.95\textwidth]{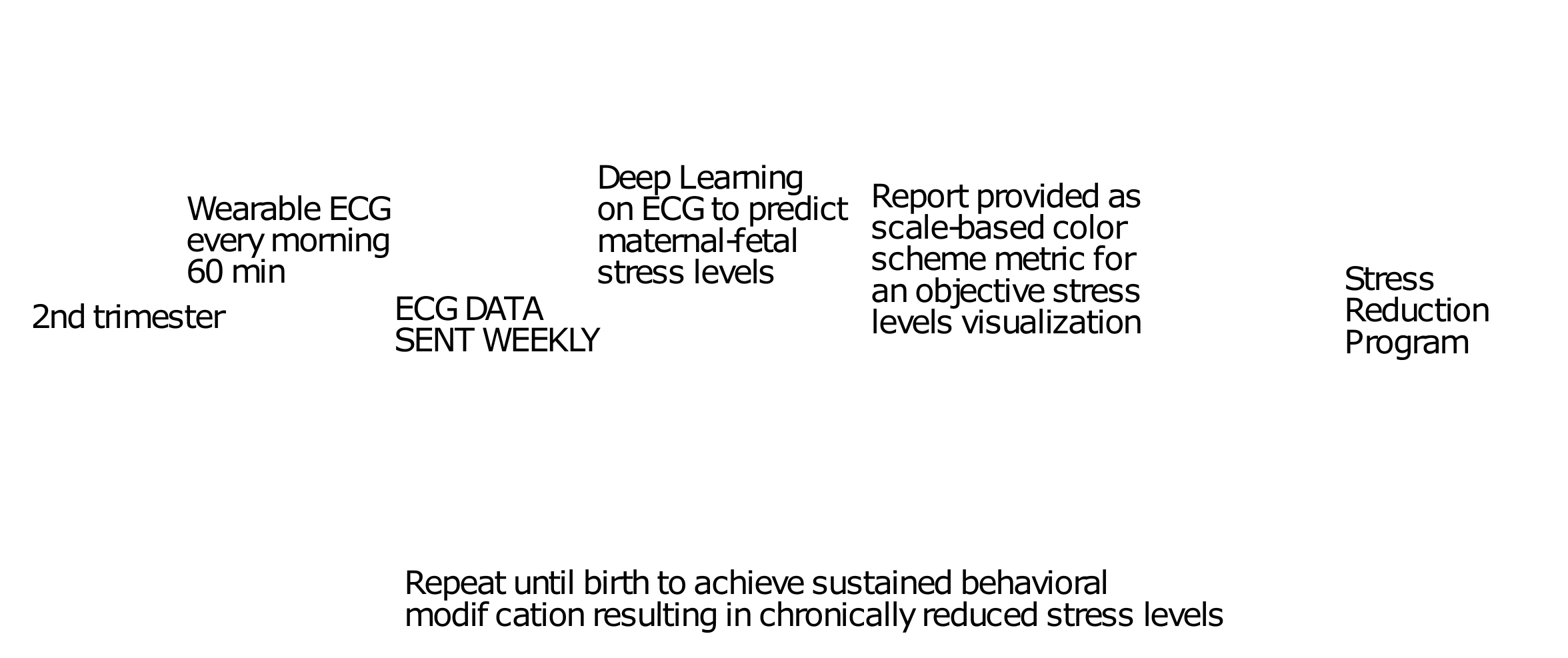}
    \caption{Real-world application of our AI model to reduce stress during pregnancy and prevent its long-term sequelae.}
    \label{innovation}
\end{figure}

This is an important advance in early and non-invasive detection of chronic stress effects during pregnancy. The demonstration of mECG being sufficient translates into the ability of using conventional ECG devices which are widely available already. This will also enable wider utilization of ECG for studies of chronic stress effects on maternal, fetal, and postnatal health. Figure \ref{innovation} demonstrates a possible deployment scenario made feasible by this work.

Another novel insight stems from two related observations. First, there was a high degree of accuracy in predicting individual characteristics of the mother-fetus dyads related to chronic stress (cortisol, FSI, PDQ, and PSS). Second, an exploration of the neural network’s latent space features suggests strongly that the entire ECG waveform structure is required and not only the temporal features of R-R peaks, i.e., heart rate variability (data not shown). 

The deep neural network properties are important to consider for two reasons. First, there appears to be a rich intrinsic integrated information about these distinct physiological properties contained in ECG. This information is retained after the temporal order is destroyed by permutation of ECG waveforms as done in this work. To our knowledge, this is the first demonstration of such a relationship. Second, most presently available wearables do not record continuous ECG, but, rather, use photoplethysmography (PPG) sensors to track heart rate triggered from the pulse waveform. A new DL approach suggests that higher quality ECG signal can be derived from PPG using a generative adversarial network (GAN) architecture \cite{sarkar2020cardiogan}. Further research is needed to validate whether this may pave the way to using the present day wearables for identifying stress. Meanwhile, a next generation of wearables is capable of continuous on-body ECG monitoring \cite{jeong2020continuous}, while some readily available clinical-grade ECG trackers can be deployed for this purpose already \cite{herry2018heart}.

Our study has limitations. First, the datasets are relatively small with less than 200 subjects in both the public and FELICITy datasets, yet the model performance has shown satisfactory stability. Also, these datasets are the largest known to us so far to permit such investigation. Second, we abstained from complicating our model with the addition of ancillary features such as BMI. It has been suggested that a bias may be added with such an approach that results from introducing unintended confounders in the causal inference sense, e.g., BMI may interact with ECG features or other features that interact with ECG in ways we don’t know and the results may be biased or even meaningless as a result. 

In conclusion, maternal-fetal early-life stress and its molecular and biophysical characteristics can be predicted with very good accuracy and reproducibility from regular ECG using a scalable SSL deep learning approach.

\section{Methods}
\subsection{FELICITy Study}
\begin{figure}[ht]
    \includegraphics[width=1\textwidth]{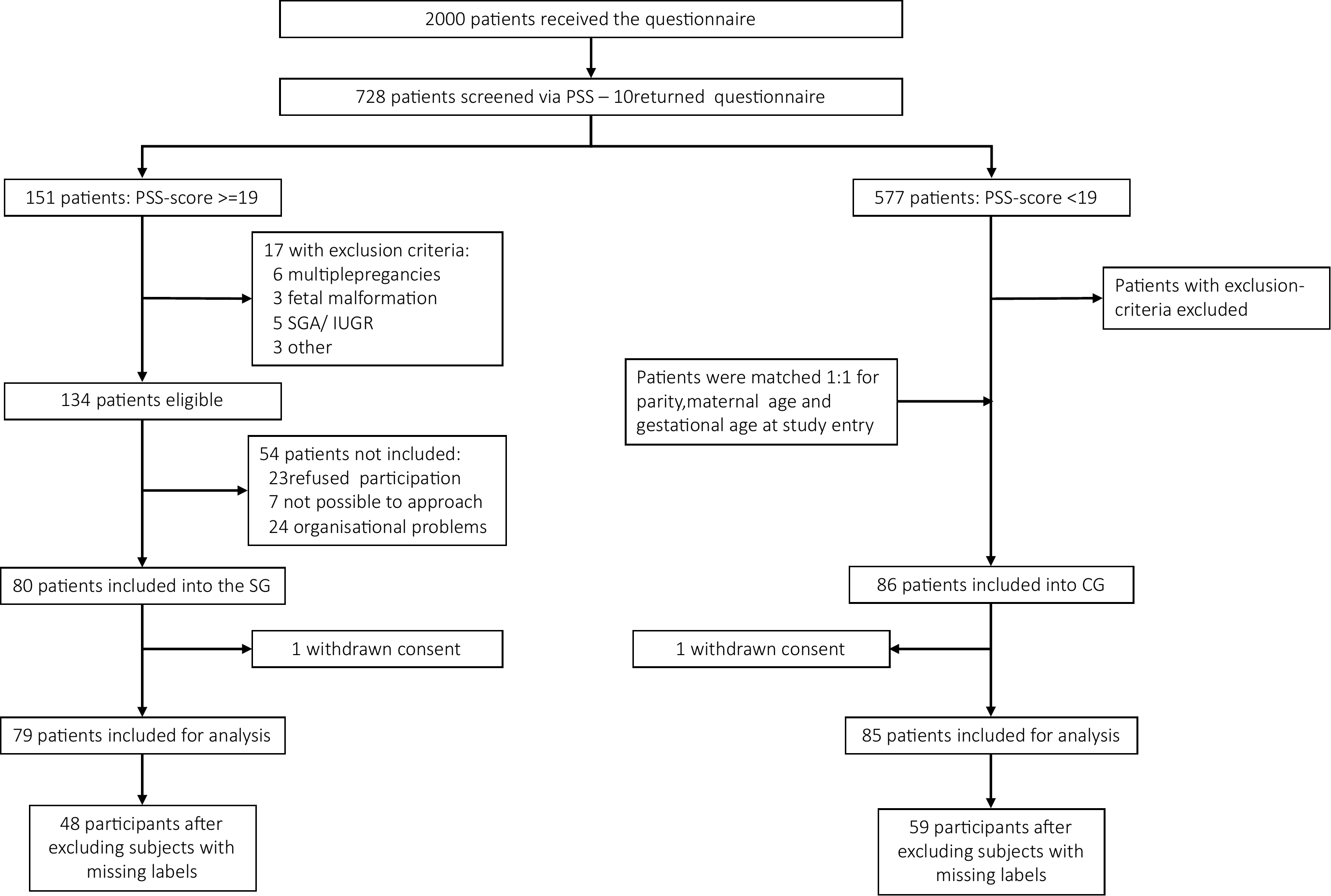}
    \caption{Recruitment flow chart for the FELICITy dataset: from screening to deep learning.}
    \label{participant}
\end{figure}

The complete experimental design can be found in \cite{lobmaier2020fetal}. Ethics approval was obtained from the Committee of Ethical Principles for Medical Research at the TUM (registration number 151/16S; ClinicalTrials.gov registration number NCT03389178). Briefly, in this prospective study, stressed mothers were matched with controls 1:1 for parity, maternal age, and gestational age at study entry. Recruited subjects were between 18 and 45 years of age, and were in their third trimester. The study ran for 22 months from July 2016 until May 2018, and subjects were selected from a cohort of pregnant women followed in the Department of Obstetrics and Gynecology at ``Klinikum rechts der Isar'' of the Technical University of Munich (TUM). This is a tertiary center of Perinatology located in Munich, Germany, which serves ~2000 mothers/newborns per year. Figure \ref{participant} presents the recruitment flowchart for this dataset and the use of data in this study. Four exclusion criteria were applied, namely (a) serious placental alterations defined as fetal growth restriction according to Gordijn et al. \cite{gordijn2016consensus}; (b) fetal malformations; (c) maternal severe illness during pregnancy; (d) maternal drug or alcohol abuse. 

The Cohen Perceived Stress Scale questionnaire was administered to gauge chronic non-specific stress exposure (PSS-10) \cite{cohen1983global}. PSS-10 $\geq19$ categorized subjects as stressed, as established \cite{lobmaier2020fetal}. We applied inclusion- and exclusion criteria following returning the questionnaires. When a subject was categorized as stressed, the next screened participant matching for gestational age at recording with a PSS-10 score $<$19 was entered into the study as control. 
In addition to PSS-10, the participants received the German Version of the “Prenatal Distress Questionnaire” (PDQ) containing 12 questions on pregnancy related fears and worries regarding pregnancy related changes of the body weight and troubles, child’s health, delivery and pregnancy’s impact on the women’s relationship.

A transabdominal ECG (aECG) recording with a sampling rate of 900 \textit{Hz} and a duration of at least 40 minutes was performed two and a half weeks after screening. The AN24 (GE HC/Monica Health Care, Nottingham, UK) was used. We calculated the signal quality index (SQI) \cite{li2017efficient} for aECG, in one-second windows, and subsequently discarded segments with an SQI of lower than 0.5. Using the fetal and maternal ECG deconvolution algorithm SAVER \cite{li2017efficient}, we extracted fetal ECG (fECG) and maternal ECG (mECG).

We utilized SQI to discard the noisy data resulting in the averaged duration of mECG and aECG of 46.07$\pm$8.74 minutes, whereas the average duration of fECG was 3.25$\pm$7.83 minutes. Due to its short duration, we could not utilize the extracted fECG data to train our self-supervised model and continued with aECG and mECG signals only. However, please note that aECG signal does represent a composite signal containing fECG and mECG, so DL on this signal tells us something about the fECG features. We refer to the resulting ECG dataset containing aECG and mECG as the FELICITy dataset.

We detected the fetal R-peaks and the maternal R-peaks separately from the respective fECG and mECG signals. The fetal- and maternal RR interval time-series were subsequently derived from the fetal and maternal R-peaks. We then calculated the mean fetal heart rate (fHR) and mean maternal heart rate (mHR) values.

Upon delivery of the baby, we recorded the clinical data including birth weight, length, and head circumference, pH, and Apgar score. Maternal cortisol was assessed using established methodology \cite{cooper2012society,iglesias2015hair,gonzalez2019hair}.

\subsection{Bivariate Phase Rectified Signal Averaging}
% The bivariate PRSA method was used to compute the Fetal Stress Index (FSI) as reported.17,18,19,1 \cite{lobmaier2012phase,lobmaier2016phase} \cite{bauer2009bivariate} FSI quantifies the entrainment between two signals, mHR as the trigger signal and the fHR as the target signal.

To analyze the relationship between two signals recorded synchronously, mHR and fHR, we use the bivariate phase rectified signal averaging (BPRSA) method \cite{bauer2009bivariate}. This method extends the ``monovariate'' PRSA method proposed for the analysis of fHR \cite{lobmaier2012phase,lobmaier2016phase}.

The two signals in question in this study are the mHR (trigger signal) and the fHR (target signal). The BPRSA algorithm operates by first detecting a number of anchor points $A$, defined as decreases in mHR. Next, for the detected set of $A$, we interpolate the fHR with a sampling rate of 900 \textit{Hz} to match the maternal ECG. We then detect the time of the anchor points in fHR, which we denote by $A'$. Then, around each anchor point $A'$ in fHR, a window of length $(2L)$ is selected. In this paper, we set $L =$ 9000, resulting in a window of 20 seconds. Next, by aligning the anchor points, we obtain phase-rectified segments. The resultant segments are then averaged to obtain BPRSA signal $X$. Consequently, we can interpret defections in $X$ as coupling between mHR and fHR. Lastly, $X$ is quantified within specific windows before and after the center of $X$. Accordingly, the designated windows are characterized as $L+S1$ to $L+S2$, and $L−S2$ to $L−S1$, where $S1$ and $S2$ are the indices used for this quantification step. We set $S1 =$ 1350 and $S2 =$ 2250, which results in windows of 1.5 and 2.5 seconds, given our sampling rate of $900$ \textit{Hz}.

Fetal stress index (FSI) is a parameter defined to analyze the coupling between mHR and fHR using the BPRSA. This index is defined as the difference between the means of the two windows mentioned above, as follows:
\begin{equation}
    FSI = \frac{1}{S2-S1} \sum_{i=L+S1}^{L+S2}X(i) - \frac{1}{S2-S1} \sum_{i=L-S2}^{L-S1} X(i),
\end{equation}
where index $L$ at the center of $X$ corresponds to our anchor definition (within the maternal RR intervals). Accordingly, the response of the fetus on mHR decreases is measured by FSI.

\subsection{Representation Learning}
\begin{figure}[ht]
    \includegraphics[width=0.7\textwidth]{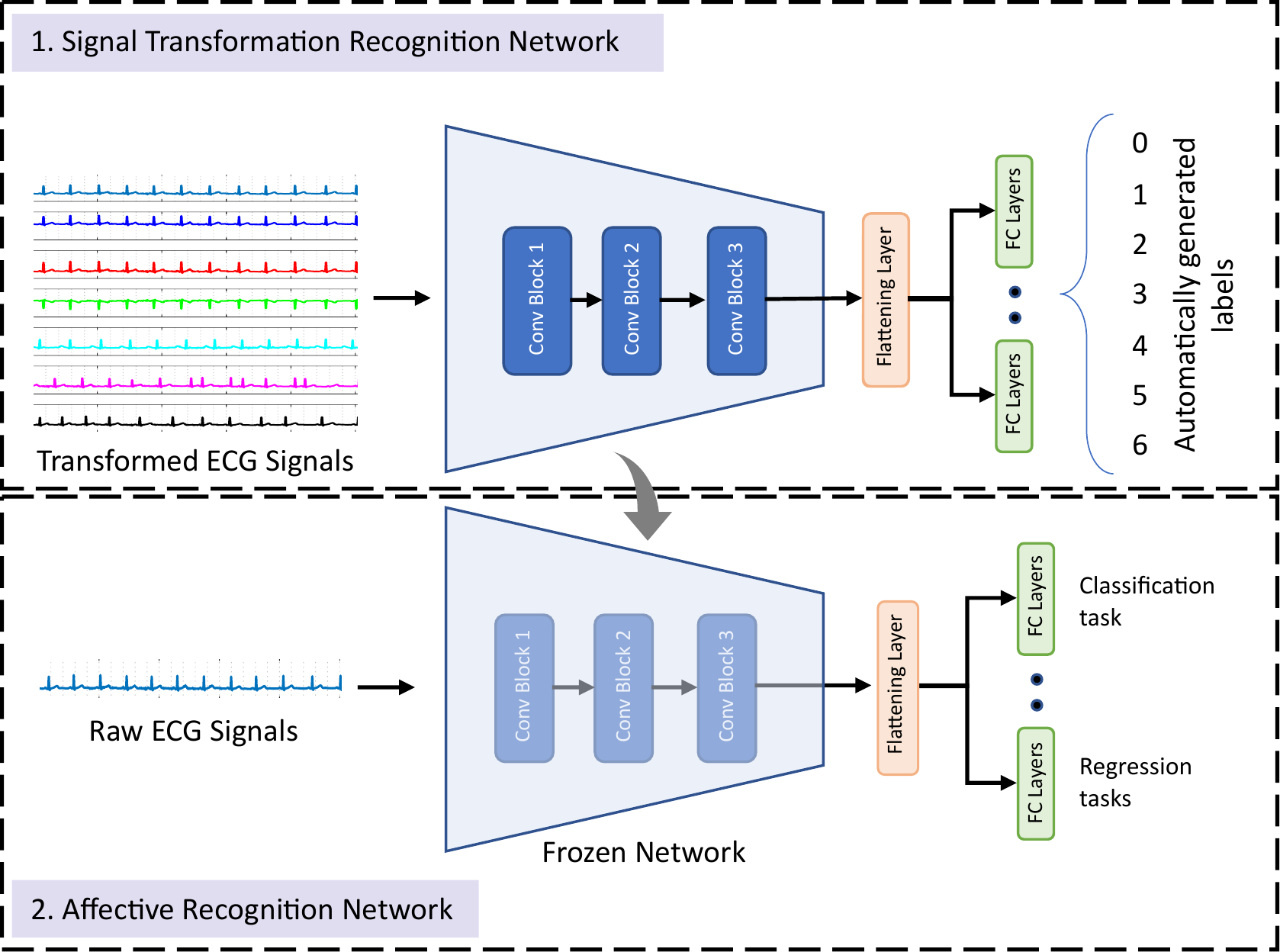}
    \caption{Our deep learning approach using a self-supervised learning framework.}
    \label{ssl_v2}
\end{figure}

We utilized an established self-supervised learning framework \cite{sarkar2020selficassp,sarkar2020selftaffc} to learn robust representations from our collected ECG data, which were further used to classify the level of stress, as well as to perform regression analyses. The framework consisted of 2 stages of learning, the first stage consisted of learning ECG representations and the second stage consisted of learning affect attributes from the learned representations (see Figure \ref{ssl_v2}). 

% \subsubsection{Convolutional Neural Network}
% We used a convolutional neural network (CNN) for learning both ECG representations and affect. The architecture of...

\subsubsection{Learning ECG Representations}
We utilized a multi-task convolutional architecture, henceforth referred to as the `transformation recognition network', which consists of 3 convolutional blocks. Each block consists of two 1D convolution layers with leaky rectified linear unit (ReLU) activation functions, followed by a max pooling layer. Following the convolutional layers, a global max pooling is used. This is finally followed by the several parallel fully connected (FC) layers.
% which was further fed to the several parallel blocks of the fully connected layers. 
We applied dropouts to reduce the possibility of overfitting. 
% We used each block of fully connected layers to perform each transformation recognition task. 
A detailed description of this network’s architecture is given in supplementary material.

In order to learn the ECG representation, the model was trained in a self-supervised manner. Automatic labels were generated through the following transformations:
\begin{enumerate}
\item Noise addition: Random Gaussian noise is added to the raw ECG signal.
\item Scaling: The magnitude of the original ECG is scaled.
\item Negation: The original ECG signal is flipped vertically.
\item Temporal Inversion: The original ECG signal is flipped horizontally.
\item Permutation: The raw ECG signal is first divided into smaller segments of equal length, which are then randomly shuffled across the time axis.
\item Time-warping: ECG signals are first divided into smaller segments similar to the permutation operation, These segments are then stretched or squeezed across the time axis. 
\end{enumerate}

The parameters of the above-mentioned transformations were derived from our previous work \cite{sarkar2020selftaffc}. Next the transformed signals were stacked randomly to create the input matrix for the self-supervised network, while the corresponding labels of the transformations were stacked, in a similar order to the inputs, to create the output labels. Each of these transformation labels are used as an output to one of the FC layers to construct a multi-task network.

\subsubsection{Learning Affect}
% Once the previous stage is complete and effective ECG representations are learned, the convo
In the second stage, affective attributes were learned using the learned ECG representations obtained from the self-supervised network. In this stage we classified stress followed by regression analysis of maternal hair cortisol, FSI, PDQ, and PSS values. The affect recognition network contains the similar convolutional layers as those used in the self-supervised network, followed by fully connected layers. The weights of the convolution layers are transferred from the signal transformation recognition network and kept frozen, and only the fully connected layers are trained. Detailed descriptions of the architectures are mentioned in the supplementary material.
% We utilized 2 fully connected layers for classification and 4 fully connected layers for the regression tasks.  

\subsection{Training}
\subsubsection{Approaches}
In order to explore the generalizability of the self-supervised method, we tackled this task in two different ways. Our first approach was to use FELICITy dataset and train the framework from scratch. As the second approach, we utilized four publicly available datasets to train the signal transformation network for learning ECG representations, followed by using FELICITy dataset to perform affect recognition by training the fully connected layers of the second network. The details of these two approaches are mentioned below. 

\noindent\textbf{First Approach - Learning From FELICITy dataset:}
As mentioned above, in our first approach, we utilized FELICITy dataset to train the self-supervised network consisting of both the signal transformation recognition network responsible for learning to extract ECG representations, as well as the fully connected layers of the affect recognition network. 
% First, we perform 6 different signal transformations and generate corresponding labels. Next, the original signal along with the transformed signals were used to train the multi-task self-supervised network. Upon successful self-supervised training, the learned feature representations along with human annotated labels were used to perform the affect recognition tasks.  

\noindent\textbf{Second Approach - Transfer Learning From Public Datasets:}
In order to explore the generalizability of the self-supervised learning, we used 4 publicly available datasets namely, AMIGOS \cite{correa2018amigos}, DREAMER \cite{katsigiannis2017dreamer}, SWELL \cite{koldijk2014swell}, and WESAD \cite{schmidt2018introducing} to train the signal transformation recognition network, i.e., learn ECG representations. Next, we transferred the weights of the network to the affect recognition network where we utilized FELICITy dataset and collected labels to train the fully connected layers of the network so that stress can be classified and factors such as maternal hair cortisol, FSI, PDQ, and PSS values can be regressed. A brief description of the public datasets is provided in supplementary material.

\subsubsection{Implementation Details}
We performed minimal pre-processing on the raw data. We re-sampled ECG signals to a sampling frequency to 256 \textit{Hz}, followed by segmentation into 10-second windows as proposed by \cite{sarkar2020selftaffc}. Next, to remove the noisy parts of aECG and mECG data, we utilized the SQI values available with the segments. To this end, SQI $<$ 0.5 were discarded. This resulted in removing approximately 4.1\% of total acquired data with a standard deviation of 8.8. In other words, approximately 50 minutes (46.07$\pm$8.74) of ECG data from each participant were used. 

We divided our whole dataset into training and test sets using a 5-fold cross-validation technique as follows. To create the training and test sets, we randomly divided each person’s data into 5 equal parts, where 4 parts were selected for training, and the 1 part was used for testing. The process was repeated 5 times.
%The same process was performed for all the subjects and repeated 5 times. 

The hyper-parameters for the self-supervised model are the same as those use in our earlier work \cite{sarkar2020selftaffc}. An Adam optimizer was used to train the models with a learning rate of 0.001 and a batch size of 128. A binary cross-entropy loss was used for the classification task and mean absolute error loss was used for the regression tasks.

The fetal and maternal ECG and HR extraction algorithms were carried out in Matlab R2016a. The SSL DL pipeline was implemented using NVIDIA GeForce RTX 2070 GPU in TensorFlow 1.14, and is publicly available at: \url{https://code.engineering.queensu.ca/17ps21/ssl-ecg-v2}.

\subsection{Statistical Analyses}
We used the Shapiro–Wilk test to evaluate for normal distribution. Medians and interquartile ranges were reported for skewed distributions, while the means and standard deviations are reported for Gaussian distributions. Where data are categorical, we present the absolute and relative frequencies. Groups are compared using t-test for independent samples, Mann–Whitney U test, and Pearson Chi-squared test.

All of the statistical tests were performed two-sided with statistical significance considered at p $<$ 0.05. The Bonferroni-Holm correction was used to adjust for multiple comparisons. To estimate the predictive performance of the quantitative variables for the presence of PS, receiver operating characteristics (ROC) analyses were carried out. Linear regression analyses were conducted to quantify the model performance in the regression tasks, expressed as R2 (see Table \ref{tab:tab3}), Mean Average Error (MAE), and Root Mean Squared Error (RMSE) (see Table \ref{tab:tab_supp}). All analyses were done with Python v3.6 Scipy library.

\section*{Acknowledgements}
We gratefully acknowledge the contribution of Dr. Hau-Tieng Wu lab with SAVER code-driven mECG/fECG extraction. The project was developed and performed by own resources of Frauenklinik/Klinikum rechts der Isar and funding from Hans Fischer Senior Fellowship to MCA.

\section*{Disclosures}
MGF has a patent pending on aECG signal separation (WO2018160890). 
% AE has patent pending on sensing biometric data (US20200314184).

\bibliography{library}
\bibliographystyle{unsrt}

\newpage
\section*{Supplementary Materials}
% \subsection*{Implementation}

% Statistical analysis was performed using IBM SPSS Statistics for Windows, version 25 (IBM Corp., Armonk, NY, USA).

\subsection*{Network architectures}
We utilize a popular convention to describe the CNN architectures. For example, cks2-f denotes a convolution layer with kernel size 1 $\times$k, stride 2, and f number of filters. mp8-s2 denotes a max-pool layer with a filter size of 8 and a stride 2. fcN denotes a fully connected layer with N hidden nodes. We utilize leaky-ReLU activation functions in all the convolution and fully connected layers, except the last layers, where sigmoid activation functions are used for the classification and recognition networks, and direct logits are extracted during the regression tasks. Finally, P$\times$[fcN] indicates P number of parallel branches in the multi-task networks. Using this convention, the details of our models are given below.  

\noindent Signal Transformation Recognition Network: 
\begin{sloppypar}{\fontfamily{qcr}\selectfont
c32s2-32, c32s2-32, mp8-s2, c16s2-64, c16s2-64, mp8-s2, c8s2-128, c8s2-128, global-max-pool, 7 $\times$  [fc128-dropout, fc128-dropout, fc1].
}\end{sloppypar}

\noindent Affect Recognition Network (classification):
\begin{sloppypar}{\fontfamily{qcr}\selectfont
c32s2-32, c32s2-32, mp8-s2, c16s2-64, c16s2-64, mp8-s2, c8s2-128, c8s2-128, global-max-pool, fc512, fc512, fc1.
}\end{sloppypar}
\noindent Affect Recognition Network (regression):
\begin{sloppypar}{\fontfamily{qcr}\selectfont
c32s2-32, c32s2-32, mp8-s2, c16s2-64, c16s2-64, mp8-s2, c8s2-128, c8s2-128, global-max-pool, 4 $\times$ [fc512, fc512, fc512, fc512, fc1]}\end{sloppypar}

\subsection*{Prediction of stress biomarkers: DL regression task}
In addition to Table \ref{tab:tab3}, we further calculate mean absolute error (MAE) and root mean square error (RMSE) of our self-supervised framework in predicting cortisol, FSI, PDQ, and PSS (Table 4).

\begin{table}[!ht]
% \fontsize{9pt}{10pt}\selectfont
\setlength\tabcolsep{3pt}
\caption{MAE and RMSE values for prediction of biomarkers using self-supervised learning, on the FELICITy and public datasets.}
\label{tab:tab_supp}

\begin{threeparttable}
\begin{tabular}{l|l|ll|ll}
\hline
% \multirow{2}{l}{Heading 1}
\multirow{2}{*}{Task} & \multirow{2}{*}{Source} & \multicolumn{2}{c|}{FELICITy dataset} & \multicolumn{2}{c}{Public datasets} \\ \cline{3-6}
& & MAE & RMSE & MAE & RMSE \\ \hline\hline
\multirow{2}{*}{Cortisol}
 & aECG & $37.832 \pm 3.284$ & $66.627 \pm 3.209$ & $16.081\pm 0.542^*$ & $40.298\pm 0.903^*$ \\ 
 & mECG & $16.445 \pm 16.068$ & $40.621 \pm 23.662$ & $6.676\pm 0.334^{\#*}$ & $23.729\pm 0.977^{\#*}$ \\\hline
\multirow{2}{*}{FSI}
 & aECG & $0.307 \pm  0.025$ & $0.541 \pm 0.022$ & $0.114\pm 0.005^*$ & $ 0.326\pm 0.013^*$ \\
 & mECG & $0.111 \pm 0.138$ & $0.282 \pm 0.164$  & $0.027\pm 0.004^{\#*}$ & $0.157\pm 0.018^{\#*}$ \\\hline
\multirow{2}{*}{PDQ}
 & aECG & $3.215 \pm 0.316$ & $5.648 \pm 0.290$ & $1.158\pm 0.057^*$ & $3.436\pm 0.151^*$ \\
 & mECG & $1.189 \pm 1.624$ & $2.873 \pm 1.984$ & $0.277\pm 0.045^{\#*}$ & $1.438\pm 0.187^{\#*}$ \\\hline
\multirow{2}{*}{PSS}
 & aECG & $3.851 \pm 0.410$ & $6.307 \pm 0.335$ & $1.409\pm 0.047^*$ & $3.808\pm 0.093^*$ \\
 & mECG & $1.438 \pm 1.772$ & $3.177 \pm 2.016$ & $0.423\pm 0.039^{\#*}$ & $1.851\pm 0.161^{\#*}$ \\\hline
\end{tabular}
\begin{tablenotes}
\item[$^*$] Public versus FELICITy dataset, Mann Whitney U test.
\item[$^\#$] mECG versus aECG within the same dataset, Mann Whitney U test.
\item[Statistical significance at p $<$ $0.025$ accounting for two comparisons (using Bonferroni-Holm correction).]
\end{tablenotes}
\end{threeparttable}
\end{table}

% \begin{figure}[t]
%     \centering
%     \includegraphics[width=0.5\linewidth]{mECG Public Datasets.pdf}
%     \caption{}
% \end{figure}

% \begin{figure*}[t]
%         \centering
%         \begin{subfigure}[b]{0.475\textwidth}
%             \centering
%             \includegraphics[width=\textwidth]{aECG FELICITy Dataset.jpeg}
%             %\caption[Network2]%
%             %{{\small Network 1}}    
%             \label{fig:mean and std of net14}
%         \end{subfigure}
%         \hfill
%         \begin{subfigure}[b]{0.475\textwidth}  
%             \centering 
%             \includegraphics[width=\textwidth]{aECG Public Datasets.jpeg}
%             %\caption[]%
%             %{{\small Network 2}}    
%             \label{fig:mean and std of net24}
%         \end{subfigure}
%         \vskip\baselineskip
%         \begin{subfigure}[b]{0.475\textwidth}   
%             \centering 
%             \includegraphics[width=\textwidth]{mECG FELICITy Dataset.jpeg}
%             %\caption[]%
%             %{{\small Network 3}}    
%             \label{fig:mean and std of net34}
%         \end{subfigure}
%         \hfill
%         \begin{subfigure}[b]{0.475\textwidth}   
%             \centering 
%             \includegraphics[width=\textwidth]{mECG Public Datasets.jpeg}
%             %\caption[]%
%             %{{\small Network 4}}    
%             \label{fig:mean and std of net44}
%         \end{subfigure}
%         \caption{The regression plots for Cortisol, FSI, PSS and PDQ are presented here while trained on both Felicity Dataset and Public Datasets. } 
%         \label{fig:reg_plot}
%     \end{figure*}

\subsection*{Description of Public Datasets}
The key metrics of each dataset (AMIGOS \cite{correa2018amigos}, DREAMER \cite{katsigiannis2017dreamer}, SWELL \cite{koldijk2014swell}, and WESAD \cite{schmidt2018introducing}) are summarized in Table \ref{tab:tab1} and are outlined in more detail below. It should be noted that all the public datasets contain ECG data and corresponding emotional ground truth labels. However, the emotional labels were not used in this study given the use of our self-supervised approach with automatically generated labels.
 
\subsubsection*{AMIGOS \cite{correa2018amigos}:}
This dataset comprises ECG and emotional labels from 40 participants. Participants were asked to watch different video clips (total 16) in order to elicit their emotional states. Shimmer ECG sensors \cite{shimmer} were used to record ECG at a sampling rate of 256 \textit{Hz}. Finally, subjective arousal and valence scores were recorded on a scale of 1 to 9 at the end of each session. 

\subsubsection*{DREAMER \cite{katsigiannis2017dreamer}:}
The DREAMER dataset comprises data from 23 participants. The emotional responses were elicited by watching emotional video clips. The clips induced different emotions such as amusement, calmness, anger, excitement, disgust among others. Similar to AMIGOS, DREAMER was also collected using Shimmer ECG sensors \cite{shimmer} at a sampling rate of 256 \textit{Hz}. At the end of each session Self-Assessment Manikins (SAM) were used to record arousal and valence scores on a scale of 1 to 5. 

\subsubsection*{SWELL \cite{koldijk2014swell}:}
25 participants comprised this dataset, where ECG data and affect scores were collected as participants performed different day-to-day office jobs, for example preparing reports, making presentations, and others. TMSI MOBI \cite{TMSi_Mobi} devices were used in this study to collect ECG signals at a sampling rate of 2048 \textit{Hz}. Finally, self-reported affect scores were collected on a scale of 1 to 9 at the end of each session. 

\subsubsection*{WESAD \cite{schmidt2018introducing}:}
17 participants comprised the WESAD dataset. A RespiBAN Professional \cite{respiban} sensor was used to collect ECG at a sampling rate of 700 \textit{Hz}. Participants went through several tasks in order to elicit their emotional states, for example, watching funny video clips during amusement condition, performing arithmetic tasks under stressed condition, reading magazines under normal conditions, and others. Finally, the Positive and Negative Affect Schedule (PANAS) scheme was used to collect emotional ground truth labels.

\end{document}